# Dynamic Behavioral Mixed-Membership Model for Large Evolving Networks

Ryan Rossi, Brian Gallagher, Jennifer Neville, and Keith Henderson

[1] Purdue University, Department of Computer Science
{rrossi, neville}@purdue.edu
[2] Lawrence Livermore National Laboratory
{bgallagher, keith}@llnl.gov

**Abstract.** The majority of real-world networks are dynamic and extremely large (e.g., Internet Traffic, Twitter, Facebook, ...). To understand the structural behavior of nodes in these large dynamic networks, it may be necessary to model the dynamics of behavioral roles representing the main connectivity patterns over time. In this paper, we propose a *dynamic behavioral mixed-membership model* (DBMM) that captures the "roles" of nodes in the graph and how they evolve over time. Unlike other node-centric models, our model is scalable for analyzing large dynamic networks. In addition, DBMM is flexible, parameter-free, has no functional form or parameterization, and is interpretable (identifies explainable patterns). The performance results indicate our approach can be applied to very large networks while the experimental results show that our model uncovers interesting patterns underlying the dynamics of these networks.

## 1 Introduction

In recent years, we have witnessed a tremendous growth in both the variety and scope of network datasets. In particular, network datasets often record the interactions and/or transactions among a set of entities—for example, personal communication (e.g., email, phone), online social network interactions (e.g., Twitter, Facebook), web traffic between servers and hosts, and router traffic among autonomous systems. A notable characteristic of these *activity* networks, is that the structure of the networks change over time (e.g., as people communicate with different friends). These temporal dynamics are key to understanding system *behavior*, thus it is critical to model and predict the network changes over time. An improved understanding of temporal patterns will facilitate for example, the development of software systems to optimally manage data flow, to detect fraud and intrusions, and to allocate resources for growth over time.

Although some recent research has focused on the analysis of dynamic networks [17,3,5,12,4,20], there has been less work on developing *models* of temporal behavior in large scale network datasets. There has been some work on modeling temporal events in large scale networks [2,28] and other work that uses temporal link patterns to improve predictive models [24]. In addition, there is work on identifying clusters in dynamic data [5,25] but these methods focus on discovering underlying communities over time—sets of nodes that are highly clustered together. In contrast, we are interested in uncovering the *behavioral* patterns of nodes in the network and modeling how those patterns change over time. The recent work on dynamic mixed-membership stochastic block models (dMMSB: [9,27]), is to our knowledge, one of the only methods suitable for modeling node-centric behavior over time. The dMMSB model identifies groups of nodes with similar patterns of linkage and characterizes how group memberships change over time. However, dMMSB assumes a specific parametric form where the groups are defined through linkage to specific nodes (i.e., in particular types of groups) rather than considering more general forms of node behavior over dynamic node sets. More importantly, the dMMSB estimation algorithm is not scalable, and therefore impractical for the types of large-scale networks analyzed in this work.

In this paper, we propose a general scalable framework for modeling node behavior that can be used for analysis of dynamic graphs where the definition of behavior can be tuned for any application. The DBMM discovers features (using the graph and attributes), extracts these features over time, and automatically learns behavioral "roles" for nodes at each timestep. Afterwards, DBMM learns a time-series of role transition probability matrices for each node. These matrices reveal how the node's local and global connectivity changes over time.

Our proposed model allows us to investigate the properties of dynamic networks and understand both global and local behaviors. The model can also be used for a variety of analysis tasks. Besides being fully automatic (no user-defined parameters) and interpretable by capturing explainable patterns, trends, and the underlying dynamical process. The main strengths of the approach includes:

- **Scalable.** The learning algorithm is linear in the number of edges in the time-interval under consideration.
- **Non-parametric and data-driven.** The model structure (i.e., number of parameters) and more generally the parameterization depends on the properties of the time-evolving network.
- **Flexible.** The definition of behavior in DBMM can be tuned for specific applications. In addition, our model can naturally incorporate attribute information into the roles. The model is also applicable for all types of time-evolving networks (e.g., transactional, bipartite, incremental, and streaming networks).

We first validate DBMM on synthetic data (§3.2), then we use it for a variety of analysis tasks: predictive modeling (§3.3), exploratory analysis (§3.4), anomaly detection (§3.5). The scalability of the approach is shown in §3.6 on networks with up to 300,000 nodes and 4 million edges—datasets that are orders of magnitude larger than could be modeled with dMMSBs.

## 2    Dynamic Behavioral Mixed-Membership Model

Our goal is to model the behavioral roles of nodes and their evolution over time. Given a sequence of network snapshots (graphs and attributes), the Dynamic

Behavioral Mixed Membership Model (DBMM) consists of (1) automatically learning a set of representative features, (2) extracting features from each graph, (3) discovering behavioral roles (4) iteratively extracting these roles from the sequence of network snapshots over time and (5) learning a predictive model of how these behaviors change over time. As an aside, let us note that DBMM is a *scalable general framework* for analyzing dynamic graphs as the model components described below can be replaced by others and each component can be appropriately tuned for any application.

### 2.1   Data Model for Temporal Networks

Given a dynamic network $\mathcal{D} = (\mathcal{N}, \mathcal{E})$, where $\mathcal{N}$ is the set of nodes and $\mathcal{E}$ is the set of edges in $\mathcal{D}$, a network snapshot $\mathcal{S}_t = (\mathcal{N}, \mathcal{E}_t)$ is a subgraph of $\mathcal{D}$ where $\mathcal{E}_t$ are the edges in $\mathcal{E}$ active at time $t$ and $\mathcal{N}_t$ are the endpoints of the edges $\mathcal{E}_t$.

### 2.2   Representing Network Behavior

The idea is to discover a set of underlying roles, which together describe the behaviors observed in the network, and then assign a probability distribution over these roles to each node in the network, which explain that node's observed behavior. Roles are extracted via a two-step process.

*Feature Discovery.* The first step is to represent each active node in a given snapshot graph $\mathcal{S}_t$ using a set of representative features. For this task, we leverage [14]. The method constructs degree and egonet measures (in/out, weighted,...), then aggregates these measures using sum/mean creating recursive features. After each aggregation step, correlated features are pruned using logarithmic binning. The aggregation proceeds recursively, until there are no new features. Formally, we discover a set of features at time $t$ denoted $\boldsymbol{V}_t$ such that $\boldsymbol{V}_t$ is an $n_t \times f$ matrix where $n_t$ is the number of active nodes and $f$ is the number of features learned from the snapshot graph $\mathcal{S}_t$. The features are extracted for each network snapshot resulting in a sequence of node-feature matrices, denoted $\boldsymbol{V} = \{\boldsymbol{V}_t : t = 1, ..., t_{max}\}$.

*Role Discovery.* The next step is to automatically discover groups of nodes (representing common patterns of behavior) based on their features. For this purpose, we use Non-negative Matrix Factorization (NMF) to extract roles [15] and extend it for a sequence of graphs. Given a sequence of node-feature matrices, we generate a rank-r approximation $\boldsymbol{G}_t \boldsymbol{F} \approx \boldsymbol{V}_t$ where each row of $\boldsymbol{G}_t \in \mathbb{R}^{n \times r}$ represents a node's membership in each role and each column of $\boldsymbol{F} \in \mathbb{R}^{r \times f}$ represents how membership of a specific role contributes to estimated feature values. For constructing the "closest" rank-r approximation we use NMF because of interpretability and efficiency, though any other method for constructing such an approximation may be used instead (SVD, spectral decomposition). More formally, given a nonnegative matrix $\boldsymbol{V}_t \in \mathbb{R}^{n_t \times f}$ and a positive integer $r < \min(n_t, f)$, find nonnegative matrices $\mathbf{G}_t \in \mathbb{R}^{n_t \times r}$ and $\mathbf{F} \in \mathbb{R}^{r \times f}$ that minimizes the functional,

$$f(\mathbf{G}_t, \mathbf{F}) = \frac{1}{2} \|\mathbf{V}_t - \mathbf{G}_t \mathbf{F}\|_F^2$$

The number of structural roles $r$ is automatically selected using Minimum Description Length (MDL) criterion. Intuitively, learning more roles, increases model complexity, but decreases the amount of errors. Conversely, learning less roles, decreases model complexity, but increases the amount of errors. In this way, MDL selects the number of behavioral roles $r$ such that the model complexity (number of bits) and model errors are balanced. Naturally, the best model minimizes, *number of bits + errors*.

We iteratively estimate the node-role memberships for each network snapshot $\boldsymbol{G} = \{\boldsymbol{G}_t : t = 1, ..., t_{max}\}$ given $\boldsymbol{F}$ and $\boldsymbol{V} = \{\boldsymbol{V}_t : t = 1, ..., t_{max}\}$ using NMF. Afterwards, we have a sequence of matrices $\boldsymbol{G}_1, \boldsymbol{G}_2, ..., \boldsymbol{G}_t, ..., \boldsymbol{G}_{t_{max}}$ where each active node at time $t$ is represented with their current role memberships.

### 2.3   Behavioral Transition Model

Given a sequence of dynamic behaviors $\boldsymbol{G} = \{\boldsymbol{G}_t : t = 1, ..., t_{max}\}$, we can learn a model of how behavior in our network changes over time. More formally, given two behavioral snapshots, $\boldsymbol{G}_{t-1}$ and $\boldsymbol{G}_t$, we learn a transition matrix $\boldsymbol{T} \in \mathbb{R}^{r \times r}$ that approximates the change in behavior from time $t-1$ to $t$. The transition matrix $\boldsymbol{T}$ represents how likely a node is to transition from role $r_i$ to role $r_j$ for that particular time interval:

$$\boldsymbol{T} = \begin{bmatrix} z^{(r_1 \to r_1)} & z^{(r_1 \to r_2)} & \cdots & z^{(r_1 \to r_m)} \\ z^{(r_2 \to r_1)} & z^{(r_2 \to r_2)} & \cdots & z^{(r_2 \to r_m)} \\ \vdots & \cdots & \ddots & \cdots \\ z^{(r_m \to r_1)} & z^{(r_m \to r_2)} & \cdots & z^{(r_m \to r_m)} \end{bmatrix}$$

where $\boldsymbol{T}$ is estimated using NMF such that $\boldsymbol{G}_{t-1} \boldsymbol{T} \approx \boldsymbol{G}_t$.

In the simple form of the model presented above, we learn $\boldsymbol{T}$ using only a single transition (i.e., $t-1$ to $t$). However, we can conceive of variations that leverage more available data by considering multiple transitions (*stacked modes*) or that smooth over a sequence of transitions using kernel functions (*summary model*). We discuss these variants in detail next.

*Stacked Transition Model.* The stacked model uses the training examples from the $k$ previous timesteps. More formally, the stacked model is defined as,

$$\begin{bmatrix} \boldsymbol{G}_{t-1} \\ \boldsymbol{G}_{t-2} \\ \vdots \\ \boldsymbol{G}_{k-1} \end{bmatrix} \boldsymbol{T} \approx \begin{bmatrix} \boldsymbol{G}_t \\ \boldsymbol{G}_{t-1} \\ \vdots \\ \boldsymbol{G}_k \end{bmatrix}$$

where $k = t - w$ and $w$ is the window size; typically $w = 10$. Let us denote the stacked behavioral snapshots as $\boldsymbol{G}_{k:t}$ where $k : t$ represents all the training examples from timestep k to timestep t.

*Summary Transition Model.* This class of models uses $k$ previous timesteps to weight the training examples at time $t$ using some kernel function. The exponential decay and linear kernels are used in this work. The temporal weights can be viewed as probabilities that a node behavior is still active at the current time step $t$, given that it was observed at time $(t - k)$. We define the summary behavioral snapshot $\boldsymbol{G}_{S(t)}$ as a weighted sum of the temporal role-memberships up to time $t$ as follows, $\boldsymbol{G}_{S(t)} = \alpha_1 \boldsymbol{G}_k + ... + \alpha_{w-1}\boldsymbol{G}_{t-1} + \alpha_w \boldsymbol{G}_t = \sum_{i=k}^{t} K(\boldsymbol{G}_i; t, \theta)$ where $\alpha$ determines the contribution of each snapshot in the summary model.

In addition to exponential and linear kernels, we experimented with the inverse linear and also tried various $\theta$ values. Overall, we found the linear kernel (and exponential) to be the most accurate with $\theta = 0.7$. Nevertheless, the optimal $\theta$ will depend on the type of dynamic network and the volatility.

*Discussion & Observations.* We have found the summary model to be the best performer for prediction tasks because of its ability to smooth over multiple timesteps. However, for precisely this reason, the summary model is more difficult to interpret. Therefore, we use the summary model for prediction tasks and the stacked representation for data analysis tasks, due to its interpretability. Let us note that to achieve better accuracy in predictions, one may also estimate local transition models for each node and use these for predicting a node's future role memberships. All of these options make our model flexible for use in a variety of applications.

We also experimented with other variants of the DBMM transition model, including a stacked-summary hybrid and mutli-state models, which make an explicit distinction between transitions from activate states and transitions from inactive states. However, we opted in favor of the simpler stacked and summary models because none of these other models provided an obvious advantage.

**Table 1.** Dataset characteristics. The number of learned features and roles provide intuition about the underlying generative process and also indicates the amount of randomness (or complexity) present in the network.

| Dataset | Features | Roles | Nodes | Edges | \|T\| | length |
|---|---|---|---|---|---|---|
| Twitter Relationships | 1325 | 12 | 310,809 | 4,095,627 | 41 | 1 day |
| Twitter (Copenhagen) | 150 | 5 | 8,581 | 27,889 | 112 | 3 hours |
| Facebook | 161 | 9 | 46,952 | 183,831 | 18 | 1 day |
| Email-Univ | 652 | 10 | 116,893 | 1,270,285 | 50 | 60 min |
| Network-Trace | 268 | 11 | 183,389 | 1,631,824 | 49 | 15 min |
| Internet AS | 30 | 2 | 37,632 | 505,772 | 28 | 3 months |
| Enron | 173 | 6 | 151 | 50,572 | 82 | 2 weeks |
| IMDB | 45 | 3 | 21,257 | 296,188 | 28 | 1 year |
| Reality | 99 | 5 | 97 | 31,694 | 46 | 1 month |

While our model currently assumes the role definitions are somewhat stationary, we have found that these roles generalize and can even be applied across different networks. Nevertheless, to remove this assumption, we could simply track the loss over time and recompute the roles when it surpasses some threshold. In the future, we plan to investigate the utility of a streaming approach using the simple modification above.

## 3    Experiments

First, we validate our model's ability to distinguish among common graph patterns using synthetic data (§3.2). Then we demonstrate DBMM on a variety of analysis tasks (§3.3–3.5). Finally, we provide performance results indicating the scalability of our model for large real-world networks (§3.6).

### 3.1    Datasets & Analysis

We apply our model using a variety of dynamic networks from different domains. See Table 1 for details. Interestingly, we find a relationship between the complexity of our behavior model and the complexity present in the graph. This is clearly shown in Table 1 by analyzing simple measures generated from our behavioral representation such as the number of learned features and the number of roles. For instance, the Internet AS topology has some hierarchical structure or recurring patterns of connectivity among ISPs and therefore our model discovers only 30 features. This is in contrast to networks with more complex patterns of connectivity such as twitter and other transaction networks like the email network. In these cases, the links are instantaneous and might only last for some duration of time, thus making more complex structures more likely.

### 3.2    Experiments on Synthetic Data

We design a graph generator to validate the ability of the DBMM to distinguish between common graph patterns. The generator constructs graphs probabilistically with four main patterns: 'center of a star', 'edge of a star', bridge nodes (connecting stars/cliques), and clique nodes. After constructing the graph, we validate our models ability to capture these patterns by measuring if the extracted features and roles represent the known probabilistic patterns. We do this by computing the pairwise euclidean distance matrix $\boldsymbol{D}$ using the initial feature matrix $\boldsymbol{V}$ (and role-membership matrix $\boldsymbol{G}$). Let $\mathbf{r}(i)$ denote the actual role (or pattern) of node $i$, then the contingency matrix is $\boldsymbol{C}[\mathbf{r}(i), \mathbf{r}(j)] = \boldsymbol{C}[\mathbf{r}(i), \mathbf{r}(j)] + \boldsymbol{D}(i,j)$

Clearly, the roles and features from nodes of the same pattern are shown to be more similar than the others (smaller values along the diagonal). See Table 2.

**Table 2.** Validating DBMM's ability to distinguish patterns. Note $C$ is row-normalized.

| Features | S-CENTER | S-EDGE | BRIDGE | CLIQUE |
|---|---|---|---|---|
| S-CENTER | **0.08** | 0.25 | 0.34 | 0.33 |
| S-EDGE | 0.27 | **0.11** | 0.25 | 0.37 |
| BRIDGE | 0.29 | 0.20 | **0.17** | 0.34 |
| CLIQUE | 0.24 | 0.24 | 0.29 | **0.23** |

| Roles | S-CENTER | S-EDGE | BRIDGE | CLIQUE |
|---|---|---|---|---|
| S-CENTER | **0.07** | 0.25 | 0.33 | 0.35 |
| S-EDGE | 0.28 | **0.10** | 0.22 | 0.40 |
| BRIDGE | 0.29 | 0.18 | **0.16** | 0.37 |
| CLIQUE | 0.24 | 0.25 | 0.29 | **0.22** |

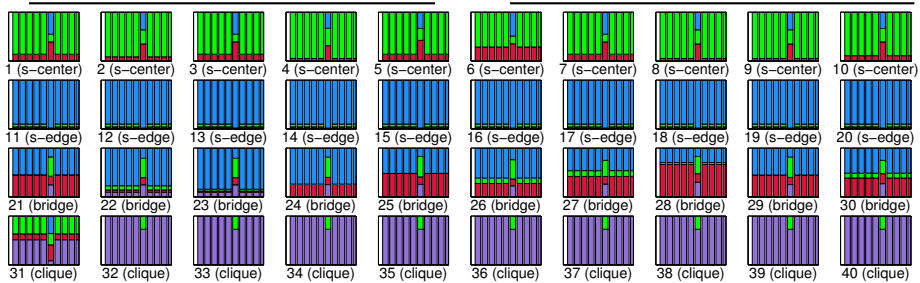

**Fig. 1.** The pattern of each node is listed below the mixed-membership plot whereas the colors represent roles learned from our model. For simplicity, the node's pattern-type is kept stable over time. Strikingly, the DBMM clearly reveals the underlying patterns of the nodes as each pattern has a distinct signature in terms of the role distribution. Moreover, the role-distributions assigned to the patterns are intuitive. For instance, the blue role of a bridge node indicates the local similarity with that of a star-edge node (low degree,...) while the red role captures the bridges more global and intrinsic property of acting as a backbone for the other nodes. The other patterns are even more straightforward to interpret. We also inject a type of global anomaly at $t = 6$ (bridges connecting to each other) which is clearly revealed as such in the plots.

Additionally, the patterns that are structurally similar to one another are represented as such by our model (star-center and clique). In Fig. 1, we visualize the mixed-memberships of 10 randomly chosen nodes from each pattern-type. Each pattern has a distinct and consistent signature in terms of the role distribution.

*Graph-based Anomaly Detection.* In a separate set of experiments, we validate our graph-based anomaly detector (see Alg. 1) by injecting anomalies into synthetic data. Initially, the dynamics of nodes are predefined to have normal transitions between patterns (e.g., star-center to clique). Then we inject some nodes with anomalous transition behavior by randomly transitioning to an abnormal pattern (star-edge to clique). For 200 repeated simulations, we achieve high accuracy (88.5%) in detecting the anomalous behavior. Using synthetic data, we have shown that DBMM can accurately recover the roles and detect anomalies.

### 3.3 Predicting Dynamic Behavior

In this section, we demonstrate the ability of DBMM to predict the future behavior of nodes. The goal is to accurately predict $G_{t+1}$ given $G_{s(t)}$, the summary behavioral snapshot described in Section 2.3. Our primary means of predicting $G_{t+1}$ is using our DBMM summary transition model $T$ as follows: $\hat{G}_{t+1} = G_t\,T$. We compare this summary model to two sensible baselines: *PrevRole* and *Avg-Role*. PrevRole simply assigns each node the role distribution from the previous timestep $t$. That is, $\hat{\boldsymbol{G}}_{t+1} = \boldsymbol{G}_t$. AvgRole assigns each node the average role distribution over all nodes at time $t$. The AvgRole model can be expressed as $\hat{\boldsymbol{G}}_{t+1} = \boldsymbol{G}_t \, \boldsymbol{T}_A$ where $\boldsymbol{T}_A$ is estimated from $\boldsymbol{G}_t = \boldsymbol{I} \cdot \boldsymbol{T}$. Essentially, PrevRole assumes node behavior does not change from timestep to timestep and AvgRole assumes that all nodes exhibit the average behavior of the network.

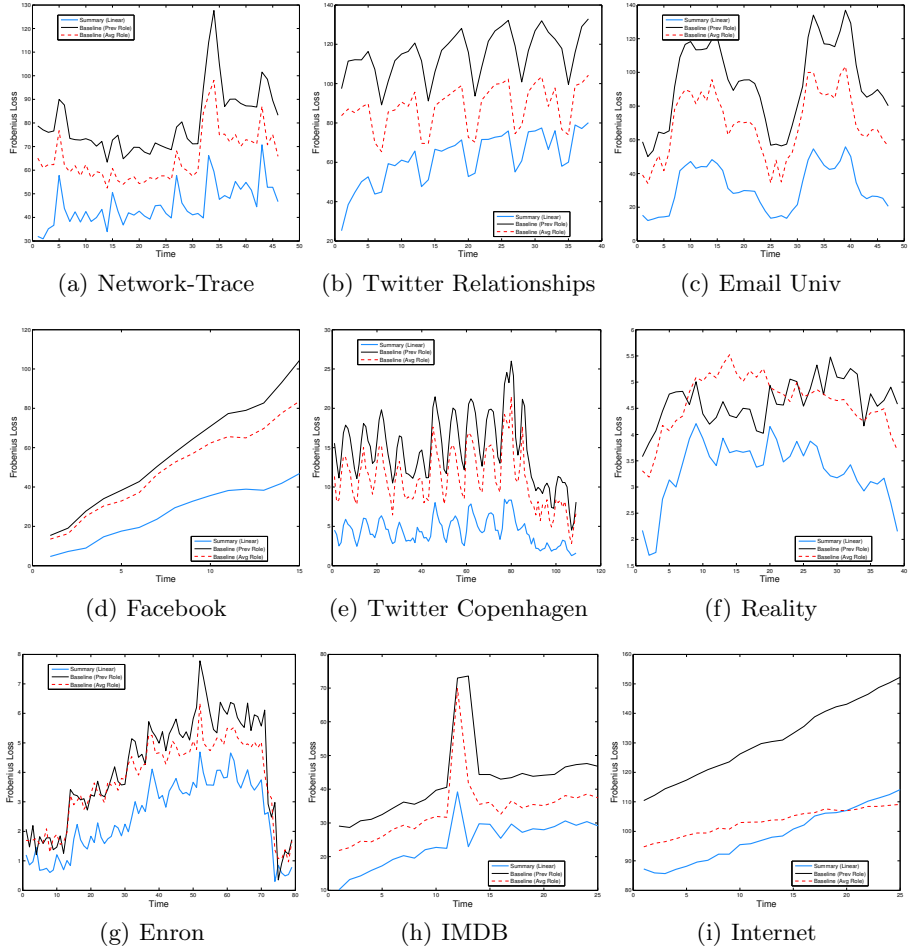

(a) Network-Trace     (b) Twitter Relationships     (c) Email Univ

(d) Facebook     (e) Twitter Copenhagen     (f) Reality

(g) Enron     (h) IMDB     (i) Internet

**Fig. 2.** The DBMM transition model accurately predicts future behavior of individual nodes (i.e., mixed role membership) compared to sensible baseline models.

We consider two strategies for evaluating our predictive models: (a) compare the predicted $\hat{\boldsymbol{G}}_{t+1}$ to the true $\boldsymbol{G}_{t+1}$ using a loss function (Frobenious norm) and (b) use $\hat{\boldsymbol{G}}_{t+1}$ to predict the modal role of each node at time $t + 1$ and evaluate these predictions using a multi-class AUC measure. We describe each of these strategies more formally below.

*Frobenious Loss.* The goal here is to estimate $\boldsymbol{G}_{t+1}$ as accurately as possible. The approximation error between the estimated node memberships $\hat{\boldsymbol{G}}_{t+1} = \boldsymbol{G}_t \boldsymbol{T}_{t+1}$ and the true node memberships $\boldsymbol{G}_{t+1}$ is defined as $||\boldsymbol{G}_{t+1} - \hat{\boldsymbol{G}}_{t+1}||_F$

*Structural Prediction with Multi-class AUC.* This is a multi-class classification task where the true class label for node $i$ is the modal role from the $i$th row of $\boldsymbol{G}_{t+1}$ (i.e., the role with max membership). The predicted class label for node $i$ is the modal role from the $i$th row of $\hat{\boldsymbol{G}}_{t+1}$. We evaluate the predictions using a generalization of AUC extended for multi-class problems (a.k.a. Total AUC) [13].

Fig. 2 shows that the DBMM summary transition model is an effective predictor across the range of experiments. With few exceptions, our model outperforms both baselines for all data sets and timesteps. This is even true for the more complex time-evolving networks such as Twitter, email, and the IP-traces, which are more transactional with rapidly evolving network structure. Note the difficulty of the prediction task varies based on the number of roles discovered, complexity of the network evolution, and the type of time-evolving network (e.g., transactional vs. social network). Some results, including the AUC plots are omitted for space, but all are qualitatively similar, see [22].

In addition to demonstrating that our model is an effective predictor, Fig. 2 offers some interesting insights into the underlying dynamics of these networks. For instance, the drift we see in Fig. 2(i) agrees with the current understanding that the underlying evolutionary process of the Internet AS is not stationary and matches recent evidence of the Internet topology transitioning from hierarchal to a flat topological structure [7,6]. Furthermore, the spike in loss (e.g., Fig. 2(h)) provide insights into network-level anomalies which could be due to large-scale emergencies, holiday seasons, or other major events.

In the larger Twitter network, we also find that users generally behave significantly different over the weekends, seen by the increase in loss and the decrease in AUC on these days (seasonality in their role transitions). Intuitively, we would expect users to be tweeting about different topics and using the system in a different manner than they do during the work days.

### 3.4 Exploratory Analysis

Our model can also uncover interesting patterns underlying the dynamics of *large-scale* networks.

*Interpretation & Analysis.* We start by applying DBMM to a large IP network (Fig. 3) where we interpret the behavior of the nodes, roles, and their evolution. In this example, we plot the evolving mixed-memberships of 4 nodes shown in Fig. 3(b) and then visualize their corresponding transition models in Fig. 3(a). The transition models are learned using the stacked representation which has shown to be better suited for exploratory analysis tasks (i.e., interpretation of the roles and their modeled transitions) whereas for prediction the summary representation yields better results.

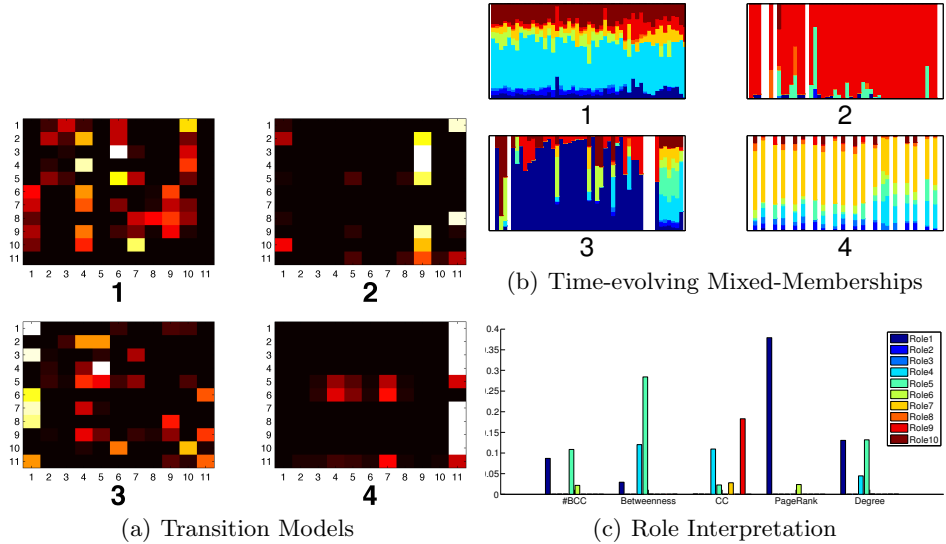

(b) Time-evolving Mixed-Memberships

(a) Transition Models

(c) Role Interpretation

**Fig. 3.** The DBMM transition model effectively captures the diverse temporal behavior of hosts in a computer network. (a) Transition matrices for 4 hosts. The y-axis represents the role the node transitions from, the x-axis is the role we transition to. Inactivity is represented by the last row/column. (b) Corresponding role-memberships over time. The x-axis represents time while the y-axis represents the role distribution at each point in time. Each distinct color represents a learned role. Inactivity is represented as white. (c) Characteristics of individual roles.

Interestingly, the evolving mixed-memberships for the nodes in Fig. 3(b) show distinct patterns from one another which are easy to identify. The temporal patterns represented by the four nodes can be classified as,

1. *Structural Stability.* This node's structural behavior (and communication pattern) is relatively stable over time.
2. *Homogeneous.* The node for the most part takes on a single behavioral role.
3. *Abrupt transition.* Their structural behavior changes abruptly. In the IP network, it could be that the IP was released and someone was assigned it or perhaps that the machine was compromised and began acting maliciously.
4. *Periodic activity.* The node has periodic activity, but maintains similar structural behavior. In the case of the IP-communication network, this machine could be infected and every 30 minutes sends out a communication to the master indicating that it is connected and "listening".

For the example nodes, we show their transition models in Fig. 3(a). The transition models represent the probability of transitioning or taking on the structural behavior of role $j$ given that your current role (or main role) is role $i$. For instance, node 2 homogeneously takes on the red role over time as discussed previously. From Fig. 3(c), we see that the red role is "role 9", and looking back at the node's learned transition model, we find that column 9 contains most of the mass, which represents that their is a high probability of transitioning from any other role to the red role. As shown in the mixed-memberships over time, this is exactly what is expected. As another example, we find that node 4 usually transitions from a mix of active roles to the inactive role (i.e., column/row eleven). Therefore, we would expect our learned transition model to capture this by placing most of the mass on the last column, representing the probability of becoming inactive after having a mix of active roles in the previous timestep, which is exactly what is found for the fourth transition model.

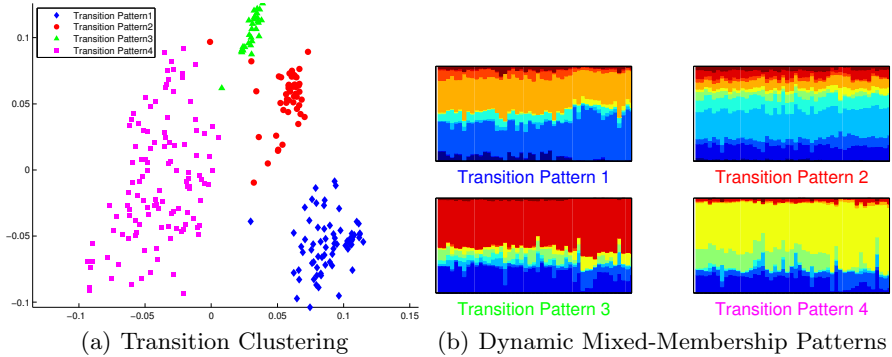

(a) Transition Clustering      (b) Dynamic Mixed-Membership Patterns

**Fig. 4.** The DBMM model provides an intuitive means of clustering nodes that exhibit similar patterns of behavior over time. (a) identifies four distinct clusters of nodes with similar transition patterns. (b) provides a sense of the behavior of each cluster in terms of the average role-membership over time.

Instead of subjective or anecdotal evidence for what the roles represent, we interpret them with respect to well-known node measures (betweenness, pagerank,...). The first technique interprets the roles using the dynamic node-role memberships $\boldsymbol{G}_t$ and a node measure matrix $\boldsymbol{M}_t \in \mathbb{R}^{n \times m}$ to compute a nonnegative matrix $\boldsymbol{E}_t$ such that $\boldsymbol{G}_t \boldsymbol{E}_t \approx \boldsymbol{M}_t$. The matrix $\boldsymbol{E}_t$ represents the contributions of the node measures to the roles at time $t$. We report average contributions over time.

Fig. 3(c) shows this quantitative interpretation of roles for the IP network. For instance, role 1 represents nodes with large PageRank, while role 5 represents nodes with large betweenness or nodes that act as bridges, whereas role 9 represents nodes with large clustering coefficient. The other roles represent more specialized structural motifs that were not captured by the set of traditional measures used for interpretation.

*Clustering Temporal Transitions.* Our model can also be used to cluster nodes based on their temporal transition patterns. This clustering reveals the underlying structural patterns of the evolving mixed-memberships. Formally, let $\boldsymbol{T}^{(i)}$ where $i = 1, ..., n$ be the estimated transition models for the nodes using the stacked model. Then we create a vector of length $r^2$ from each of the transition models and define a similarity function between these vectors. Next, we apply the classical k-means clustering algorithm to cluster the nodes by their transition matrices. Afterwards, we compute the closest rank-k approximation (k = 2 or 3) of the input matrix. The nodes are plotted using the low-rank approximation and labeled using the clustering algorithm. To reveal the structural transition pattern, we compute the average dynamic mixed-membership for each cluster.

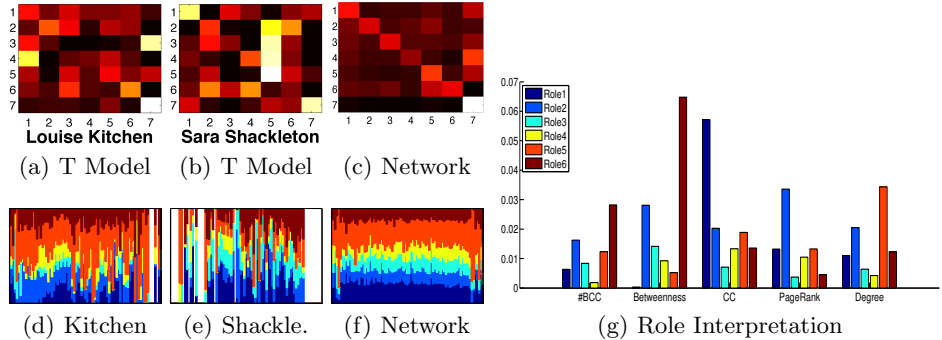

**Fig. 5.** The DBMM transition model provides an effective means of automatically discovering and visualizing nodes with anomalous temporal behavior. (a)–(b) are the transition models for two of the most anomalous nodes in the Enron email network compared to (c) the normal network transition model. (d)–(f) show the corresponding role memberships over time. (g) shows the characteristics of roles.

This clustering method reveals common structural trends and patterns between nodes. For instance, this technique groups nodes together that share similar transitional patterns such as nodes with stable roles vs. nodes with more dynamic roles or nodes with high activity vs. nodes with low activity. An example is provided in Fig. 4. For clarity in the visualization, we identified common transition patterns among only a small sample of the ≈ 180k candidate nodes. The first visualization in Fig. 4(a) identifies four distinct well-separated clusters of nodes with similar transition models. Fig. 4(b) shows the average evolving mixed-membership for each cluster. This visualization shows that each cluster represents a unique structural transition pattern between the nodes. The structural patterns can be interpreted using the previous role interpretation from Fig. 3(c). This technique can be used for general exploratory analysis such as characterizing the patterns and trends of nodes or eventually used as a means to detect anomalies or nodes that do not fit any transition pattern.

### 3.5 Detecting Graph-based Anomalies in Dynamic Networks

Previously, we validated the anomaly detection algorithm on synthetic data (see §3.2). We further demonstrate the use of DBMMs for detecting graph-based anomalies in large real-world dynamic networks. In particular, we formulate this problem with respect to identifying nodes that have unusual structural transition patterns. For instance, a node might transition from being a hub (i.e., a node with many people linking to it) to a node with high clustering coefficient.

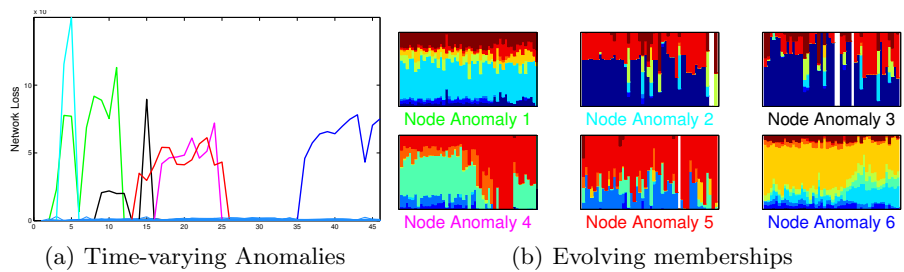

(a) Time-varying Anomalies          (b) Evolving memberships

**Fig. 6.** The DBMM model reveals nodes that are anomalous for only short periods of time and normal otherwise. Such temporally local anomalies are often impossible to find using static graph analysis because brief abnormal periods are drowned out by mostly normal behavior. (a) shows examples of short lived anomalies in a computer network. (b) shows the corresponding behavior over time for each node in detail.

*Node Anomaly.* While there are many ways to detect anomalies using the DBMM model, we propose an intuitive algorithm shown in Alg. 1 that uses a node's transition model for predicting the network memberships at $t+1$. The anomaly score is the difference between the predicted network mixed-memberships and the ground-truth mixed-memberships. Therefore, the score represents the divergence of that nodes transitions from the entire network. One simple example is shown in Fig. 5(a) where our algorithm identifies Louise Kitchen (Founder/president of EnronOnline) as having unusual behavioral transitions (large anomaly score).

*Time-varying Node Anomalies.* For detecting the specific time interval in which a node has unusual behavior we use the previous method with a few subtle distinctions. The global and node models are estimated at each *time* (in a sort of streaming fashion) using the stacked representation with a shorter window (for leveraging past training examples). The final result is a time-series $\mathbf{x}_t$ : $t = 1, ..., t_{max}$ of anomaly scores for the nodes, shown in Fig. 6. The anomalies represent nodes that transition from their normal role patterns into roles that are less likely (hub node transitioning to a bridge node). This approach not only recognizes that a node is anomalous, but captures the time-interval for which the node takes on anomalous behavior. In Fig. 6, we find that these types of time-varying anomalies occur frequently in the IP-trace network. Perhaps this is due to IP addresses (nodes) that are frequently released or expire, which are then assigned to another user who may take on significantly different behavior.

Our approach also identifies anomalies in the Twitter dataset from the cop15 UN climate change conference. For example, the users that were still actively tweeting towards the last few days of the conference become more personal with one another, forming more densely connected subgraphs. These users are anomalous since they transition from their normal behavior patterns to the more unlikely behaviors. We also find that the DBMM anomaly detector effectively captures differences in both static and dynamic behavior. These results were removed for brevity, but can be found in [22].

### 3.6 Scalability and Complexity

Most importantly, the DBMM is linear in the number of edges. The complexity is $O(|E| \cdot |T|)$ where $|T|$ is usually a trivial factor compared to the edges (even in the case where we use minute timesteps for analyzing IP-traces). A more accurate upperbound on the complexity can be stated in terms of the maximum number of edges at any given timestep. Thus, the complexity is $O(\max_t(|E|_t) \cdot |T|)$.

Our model is capable of handling realistic networks such as social and technological networks consisting of millions of nodes and edges. This is in contrast to similar dynamic mixed-membership models that have been recently proposed such as the dMMSB [27,9]. These models are quadratic in the number of nodes and therefore unable to scale to the realistic networks with the number of edges in the millions. Furthermore, these models have been typically been used for visualizing trivial sized networks of 18 nodes up to 1,000 nodes. This is in contrast to our paper where we apply DBMM not only for visualizations, but for a variety of analysis tasks using large dynamic networks.

Moreover, the dMMSB can handle 1,000 nodes in a day [27] (See page 30), while our model handles ≈8,000 nodes in 506.61 seconds (or 8 minutes and 26 seconds) shown in Table 3. We provide performance results for other larger datasets of up to 183,389 nodes and 1,631,824 edges. In all cases, even for these large networks with over a million edges, our model takes less than a day to compute and the performance results show the linearity of our model in the number of edges. For the scalability experiments, we recorded the performance results using a commodity machine Intel Core i7 @2.7Ghz with 8Gb of memory.

In addition, the proposed dynamic behavioral mixed-membership model is also trivially parallelizable (using Hadoop on Amazon EC2/Cloud) as features, roles, and transition models can be learned at each timestep independent of one another. This parallelization makes our model even more attractive for real-time analysis of large streaming graphs.

## 4  Related Work

There has been an abundance of work in analyzing dynamic networks. However, the majority of this work focuses on dynamic patterns [10,17,16,21,25], tempo-

---

**Algorithm 1** Anomalous Structural Transitions

**Input: G** $= \{\boldsymbol{G}_t : t = 1, ..., t_{max}\}$ (evolving mixed-memberships)
**Output: x** (vector of anomalous scores)

1: **for** $i = 1 \rightarrow n$ **do**
2:     $\boldsymbol{T}^{(i)} \in \mathbb{R}^{r \times r} \leftarrow NMF(\boldsymbol{G}_{1:t-1}^{(i)}, \boldsymbol{G}_{2:t}^{(i)})$
3:     $\hat{\boldsymbol{G}}_{t+1} = \boldsymbol{T}^{(i)} \cdot \boldsymbol{G}_t$
4:     $\mathbf{x}^{(i)} = \left\| \hat{\boldsymbol{G}}_{t+1} - \boldsymbol{G}_{t+1} \right\|_F$
5: **end for**

ral link prediction [8], anomaly detection [1], dynamic communities [18,26,11], dynamic node ranking [19,23], and many others [28,12].

In contrast, we propose a scalable dynamic mixed-membership model that captures the node behaviors over time and consequently learns a predictive model for how these behaviors evolve over the time. Perhaps the most related work is that of [9] where they develop the dMMSB model to discover roles in the graph and how these memberships change over time. However, this type of mixed-membership model assumes a specific parametric form, which is not scalable (1,000 nodes takes a day to model), and where the groups are defined through linkage to specific nodes (in particular types of groups) rather than more general node behavior or structural properties [27]. This is in contrast to our proposed model, which is based on our intuitive behavioral representation and can be interpreted quantitatively. In addition, our model is not tied to any single notion of behavior and thus is flexible in the roles discovered and generalizable. Moreover, not only do we evaluate our model on detecting unusual behavior, identifying explainable patterns and trends, and for clustering nodes with respect to their transition patterns, but we apply our model on large real-world networks to demonstrate its scalability. To the best of our knowledge, our proposed model is the first scalable dynamic mixed-membership model capable of identifying explainable patterns and trends on large networks.

## 5   Conclusions

We proposed a dynamic behavioral mixed-membership model that is suitable for analyzing large-scale evolving networks. The proposed model provides a general scalable framework for analysis of dynamic networks where the definition of behavior can be tuned for specific applications. We validated our model on both synthetic and large real-world networks where our model was shown to reveal interesting patterns underlying the dynamics of these networks. In future work, we plan to investigate our approach on graph streams.

**Table 3.** Performance Analysis of the Dynamic Behavioral Mixed-Membership Model. The dMMSB takes a day to handle 1,000 nodes [27], while our model takes only 8.44 minutes for 8,000 nodes.

| Dataset | Nodes | Edges | Performance |
|---|---|---|---|
| ENRON | 151 | 50,572 | 117.51 seconds |
| TWITTER (COPEN) | 8,581 | 27,889 | 506.61 seconds |
| FACEBOOK | 46,952 | 183,831 | 1,468.65 seconds |
| INTERNET AS | 37,632 | 505,772 | 1,922.85 seconds |
| NETWORK-TRACE | 183,389 | 1,631,824 | 16,138.71 seconds |